# THE COSMOLOGICAL CONSTANT FOR THE CRYSTALLINE VACUUM COSMIC SPACE MODEL

J. A. Montemayor-Aldrete[1],  J. R. Morones-Ibarra[2],  A. Morales-Mori[3],
A. Mendoza-Allende[1],  A. Montemayor-Varela[4], M. del Castillo-Mussot[1] and G.J. Vázquez[1]

1. Instituto de Física, Universidad Nacional Autónoma de México, Apartado Postal 20-364, 01000 México, D. F.

2. Facultad de Ciencias Físico-Matemáticas, Posgrado, Universidad Autónoma de Nuevo León. Apartado Postal 101-F, 66450 San Nicolás de los Garza.

3. Centro de Ciencias Físicas, Universidad Nacional Autónoma de México, Apartado Postal 139-B, 62191 Cuernavaca, Morelos. México.

4. Centro de Mantenimiento, Diagnóstico y Operación, Iberdrola, Polígono Industrial El Serrallo, Castellón de la Plana, C.P. 12100, España.

**ABSTRACT**

The value of the cosmological constant arising from a crystalline model for vacuum cosmic space with lattice parameter of the order of the neutron radius [1] has been calculated. The model allows to solve, in an easy way, the problem of the cosmological constant giving the right order of magnitude, which corresponds very well with the mean value of matter density in the universe. The obtained value is about $10^{-48}$ $Km^{-2}$. Diffraction experiments with non-thermal neutron beam in cosmic space are proposed to search for the possibility of crystalline structure of vacuum space and to measure the lattice parameter.

PACS numbers: 98.80.Es, 04.20.-q, 03.65.-w, 61.50.-f, 98.80.Ft

# I. INTRODUCTION

A review of recent cosmological observations suggests a universe that is lightweight (matter density about one third of the critical value), is spatially flat at big scale and in an unexpected way the radial distance between galaxies is increasing at an accelerating rate [1, 2].

The acceleration of the expansion of the radial distance between galaxies which forms the universe requires the existence of energy to overcome the gravitational self-attraction of matter. The cosmological constant also called lambda (written as $\lambda$ or $\Lambda$) is a long time candidate for serving as this energy reservoir. In 1967, Zel'dovich [3] showed that the energy density of the vacuum should act precisely as the energy associated with the cosmological constant. Lately, theorists have been dusting its off again and speculating about sources for the energy based on the fleeting particles that wink in and out of existence in vacuum space, according to quantum relativistic mechanics. But calculations based on that idea lead to lambda's that are 120 orders of magnitude larger than the energy contained in all the matter in the universe [4-8]. And, this last result is based on speculation, of many theorists, that there is a cutoff in the maximum frequency of the oscillation vacuum modes which corresponds to the Planck's distance of about $10^{-33}$ cm [9]. So theorists are exploring different alternatives. For instance, some researchers consider that the cosmological constant arises from different possibilities such as: Local voids or nonhomogeneities in the universe expansion [10, 11], a true Casmir effect on a scalar field filling the universe [12], or give alternative scenarios to a pure cosmological constant provided by a classical scalar field known as quintessence [2, 13-17]; also the self-tuning brane scenario like an attempt to solve the cosmological constant problem have been used [18], and some people use the anthropic principle trying to explain the small value of the cosmological constant [19, 29]. As far as we know there are no experimental data which give support in a conclusive way, to any of the previous models for the cosmological constant problem. In other words, this problem and also its associated one of the acceleration of the universe expansion remain still unsolved.

The quantum theory of the vacuum considers it as a three dimensional harmonic oscillator. Quantum field theory is notorious for its divergences. The most fundamental one concerns the energy density of the vacuum [9]. In a companion paper, the problem of gravitational stability of an infinite, three-dimensional, vacuum cosmic space with crystalline structure has been studied [21]. Such model assigns physical reality to the three dimensional harmonic oscillator scheme used in quantum theory to describe the vacuum. In this scheme the lattice parameter of the vacuum cosmic crystalline space is of the order of the neutron radius.

The main purpose of this work is to evaluate the cosmological constant arising from such crystalline vacuum cosmic space, and make a comparison with the available experimental data.

## 2. THEORY

According to our theoretical scheme [21, 22] we have the following situation. The cut off maximum frequency used in standard quantum theories implies $10^{20}$ neutron masses inside a volume $10^{60}$ time lower than the neutron volume, which gives $10^{80}$ neutron density. In our scheme, which considers that the vacuum cosmological space has a crystalline character with a lattice parameter equal to the nuclear radius of a neutron, the problem of $10^{120}$ becomes a problem of $10^{40}$ time the magnitude of the observational limits; but this of course is the expected magnitude for the case of a crystalline lattice with such lattice parameter. And it is clear that the energy density of the vacuum associated to a crystalline lattice is not a thermodynamical free energy but a linked one, which is another way to say that inside the actual universe volume, $V_{OU}$, a crystalline structure with minimum free energy exists; and such vacuum structure has minimum configurational entropy. The quantum - gravitational interaction between the lattice entities which yield the gravitational stability of such crystalline space has been described in a global way by the analysis of quantum fluctuations on the energy per unit volume of the crystalline vacuum space, $E_{CV}$, [21]. The global value of such fluctuations is about $10^{-40}$ $E_{CV}$. Therefore in our scheme the impossible problem to be solved, namely the cosmological constant problem becomes a trivial one.

In a mathematical way, we have the following, an expression which relates both the time derivatives of the cosmological constant $\frac{d\lambda}{dt}$ and the volumetric density of vacuum energy $\frac{dU_V}{dt}$. See Eq. (49) form a companion paper [22], which reads,

$$\frac{d n^+_B}{dt} = \frac{10^{80}}{R_{OU}} \frac{d\lambda}{dt} = \frac{10^{80}}{R_{OU}} \frac{8\pi G}{C^4} \frac{dU_V}{dt}, \tag{1}$$

where $R_{OU}$ is the actual universe radius, $\frac{10^{80}}{R_{OU}} \frac{d\lambda}{dt}$ the production rate of matter per unit volume, $G$ the gravitation constant and $C$ the light speed in vacuum. Equation (1) can be rewritten as [23, 24].

$$\lambda = \frac{8\pi G}{C^4} U_V \tag{2}$$

From Eq. (1) is straightforward to see that an expression for both the total number of baryons, $n_B$, and the total number of lattice entities, $n_{Le}$, inside a volume $V_{OU}$ is obtained

$$n_{Le} + n_B = \frac{10^{80}}{R_{OU}} \int_o^{V_{OU}} \int_o^{t_{OU}} \frac{d\lambda}{dt} dt dV = \frac{10^{80}}{R_{OU}} \frac{8\pi G}{C^4} \int_o^{V_{OU}} \int_o^{t_{OU}} \frac{dU_V}{dt} dt dV \tag{3}$$

In other words we have,

$$\int_o^{V_{OU}} \int_o^{V_{OU}} \frac{dU_V}{dt} dt dV = V_{OU} \{U_{VC} + U_{qf}\}, \tag{4}$$

where $U_{VC}$ is the internal energy density of the perfect crystalline vacuum cosmic space, and $U_{qf}$ is the internal energy density due to quantum fluctuations which stabilize the crystalline structure of vacuum against gravitational instabilities. With $U_{qf}$ defined as

$$U_{qf} \equiv \frac{E_{qf}}{V_{OU}}, \tag{5}$$

where $E_{qf} = E_{OU} = 10^{120} \varepsilon_{OU}$ $\qquad(6)$

and $\quad \varepsilon_{OU} \geq \frac{\eta c}{R_{OU}} \equiv h\nu_{OU},$ $\qquad(7)$

$\varepsilon_{OU}$ being the elementary quantum of gravitation energy inside a volume $V_{OU}$, which states the minimum absolute temperature $(kT_{min} \equiv kT_{OU} \geq h\nu_{OU})$ available inside the actual universe volume, $T_{OU} \cong 10^{-27} K$ [21, 22].

The total energy density of vacuum cosmic space, $U_{TV}$, could be defined as,

$$U_{TV} \equiv U_{VC} + \frac{c^4}{8\pi G} \lambda_{qf}, \tag{8}$$

where the first term is a constant ($U_{VC}$) and describes the internal energy density due to a perfect crystalline lattice of the vacuum cosmic space, with a lattice parameter close to the neutron radius; and the second term due to the long range quantum fluctuations, which stabilize the crystalline structure of vacuum against gravitational instabilities, has an energy density of $U_{qf} = 10^{-40} U_{VC}$

In other word, its possible to define the total cosmological constant, $\lambda_T$, given by

$$\lambda_T = \lambda_{VC} + \lambda_{qf} \tag{9}$$

From Eqs. (3) and (4) the following equation arises,

$$n_{Le} + n_B = \frac{10^{80}}{R_{OU}} \frac{8\pi G}{C^4} \left[V_{OU}(U_{VC} + U_{qf})\right], \tag{10}$$

which means

$$n_{Le} = \frac{10^{80}}{R_{OU}} \frac{8\pi G}{C^4} V_{OU} U_{VC}, \text{ and } n_B = \frac{10^{80}}{R_{OU}} \frac{8\pi G}{C^4} V_{OU} U_{qf} \tag{11}$$

and by using Eq. (2) these equations become,

$$n_{Le} = \frac{4\pi 10^{80} R_{OU}^2}{3} \lambda_{VC} \tag{11.a}$$

and

$$n_B = \frac{4\pi 10^{80} R_{OU}^2}{3} \lambda_{qf} \tag{11.b}$$

From Eqs. (11), the following relationship could be written

$$\frac{n_B}{n_{Le}} = \frac{\lambda_{qf}}{\lambda_{VC}} = 10^{-40} \tag{12}$$

or

$$\lambda_{qf} = 10^{-40} \lambda_{VC} \tag{13}$$

which is our expected value, $\lambda_{qf} \sim 1$ in units where $\frac{8\pi G}{C^4} U_{VC} \equiv 1$, because as implied in previous paragraphs $\lambda_{VC} \sim 10^{40}$ in such a unit system.

## 3. DISCUSSION AND CONCLUSIONS

The value for $\lambda_{qf}$ could be expressed in other, more conventional, way. From Eqs. (11.a) and (11.b), one has

$$\lambda_{qf} = \frac{3}{4\pi} \frac{1}{R_{OU}^2} \qquad (14)$$

and

$$\lambda_{VC} = \frac{3}{4\pi} \frac{10^{40}}{R_{OU}^2} \qquad (15)$$

where $n_B = 10^{80}$ and $n_{Le} = 10^{120}$.

Equation (14) gives $\lambda_{qf} \cong 10^{-48} \, K \, m^{-2}$ which is the right order of magnitude for an Euclidean character of cosmic space at large scale as actually observed.

In this theoretical scheme, it is not possible to obtain another value for $\lambda_{qf}$ than the one with the right one order of magnitude, and the problem of the cosmological constant becomes a trivial one. Also, it follows immediately that the total energy density of the vacuum space $U_{TV}$ can be locally determined from diffraction experiments by using a neutron beam in conditions of very high vacuum levels. In principle the diffraction experiments are easy to imagine, a very well collimated non - thermal neutron beam is ejected into the cosmic vacuum, some of such neutrons will be diffracted by the vacuum crystalline structure exhibiting a change on linear momentum due to their interaction with the vacuum lattice to its interaction with the vacuum lattice entities. In principle, careful diffraction experiments should allow us also to obtain information about the long - range gravitational stresses on the crystalline structure of the neighboring cosmic space.

## ACKNOWLEDGEMENTS

We want to specially thank Prof. M. López de Haro for many years of deep discussions and arguments, for his contribution to final shaping of the ideas and for his encouragement not to give up and unorthodox approach to cosmological problems and also we acknowledge to the librarian Technician G. Moreno for her stupendous work and to O. N. Rodríguez Peña for her patient and skilful typing work.